\documentclass[nopreprintline]{elsarticle}

\usepackage{graphicx}
\usepackage[caption=false,font=scriptsize]{subfig}
\usepackage{multirow}
\usepackage{multicol}
\usepackage{hyperref}
\usepackage{amssymb,array}
\usepackage{color}
\usepackage{amsmath,dsfont}
\usepackage{booktabs}
\usepackage{threeparttable}
\usepackage{textcomp}
\usepackage{url}

\usepackage{xspace}
\makeatletter
\DeclareRobustCommand\onedot{\futurelet\@let@token\@onedot}
\def\@onedot{\ifx\@let@token.\else.\null\fi\xspace}

\def\eg{\emph{e.g}\onedot} 
\def\ie{\emph{i.e}\onedot}

\def\etal{\emph{et al}\onedot}
\makeatother

\newcommand{\PreserveBackslash}[1]{\let\temp=\\#1\let\\=\temp}
\newcolumntype{C}[1]{>{\PreserveBackslash\centering}p{#1}}
\newcolumntype{R}[1]{>{\PreserveBackslash\raggedleft}p{#1}}
\newcolumntype{L}[1]{>{\PreserveBackslash\raggedright}p{#1}}

\begin{document}

\begin{frontmatter}

\title{Identification of Deep Network Generated Images Using Disparities in Color Components}

\author[mymainaddress,mysecondaryaddress]{Haodong Li}
\ead{lihaodong@szu.edu.cn}

\author[mymainaddress,mysecondaryaddress]{Bin Li\corref{mycorrespondingauthor}}
\cortext[mycorrespondingauthor]{Corresponding author}
\ead{libin@szu.edu.cn}

\author[mymainaddress,mysecondaryaddress]{Shunquan Tan}
\ead{tansq@szu.edu.cn}

\author[mymainaddress,mysecondaryaddress]{Jiwu Huang}
\ead{jwhuang@szu.edu.cn}

\address[mymainaddress]{
  Guangdong Key Laboratory of Intelligent Information Processing, Shenzhen Key
  Laboratory of Media Security, Guangdong Laboratory of Artificial Intelligence
  and Digital Economy (SZ), Shenzhen University, Shenzhen 518060, China}
\address[mysecondaryaddress]{
  Shenzhen Institute of Artificial Intelligence and Robotics for Society,
  Shenzhen 518172, China}


\begin{abstract}
With the powerful deep network architectures, such as generative adversarial
networks, one can easily generate photorealistic images. Although the generated
images are not dedicated for fooling human or deceiving biometric authentication
systems, research communities and public media have shown great concerns on the
security issues caused by these images. This paper addresses the problem
of identifying deep network generated (DNG) images. Taking the differences
between camera imaging and DNG image generation into considerations, we analyze
the disparities between DNG images and real images in different color
components. {We observe that the DNG images are more distinguishable
from real ones in the chrominance components, especially in the residual
domain}. Based on these observations, we propose a feature set to capture color
image statistics for identifying DNG images. Additionally, we evaluate several
detection situations, including the training-testing data are matched or
mismatched in image sources or generative models and detection with only real
images. Extensive experimental results show that the proposed method can
accurately identify DNG images and outperforms existing methods when the
training and testing data are mismatched. Moreover, when the GAN
model is unknown, our methods also achieves good performance with one-class
classification by using only real images for training.
\end{abstract}


\end{frontmatter}


\section{Introduction}
\label{sec:introduction}

In recent years, we have witnessed the inspiring development of image generative
models \cite{goodfellow2014generative,kingma2014auto,oord2016pixel}.
Traditionally, image generative models could only create simple image textures,
and the image contents were far from realistic. Therefore, it was not difficult
to differentiate between such generated images and real images with naked eyes. However, the
situation has changed with the tremendous advancement of machine learning. With
the modern deep network architectures, especially the generative adversarial
networks (GANs) \cite{goodfellow2014generative}, the quality of generated images
has been dramatically improved
\cite{karras2018progressive,brock2018large,karras2018style}, and thus it is no
longer easy to identify the deep network generated (DNG) images with human
visual system. Although the advancements of image generative models have
facilitated many applications, such as image super-resolution
\cite{ledig2017photo,wang2019variational}, image translation
\cite{zhu2017unpaired,liu2017unsupervised}, and image inpainting
\cite{iizuka2017globally}, the photorealistic DNG images may also lead to many
serious security risks. For example, the generated scenes can be used to falsify
images or videos and fabricate fake news, the generated faces can be posted on
social networks to counterfeit personal information or be used to attack
biometric authentication systems. Recently, both public media \cite{snow2017ai}
and research communities \cite{knight2018us} have shown great concerns on the
negative impacts of DNG images, and some governments \cite{2019virginia} have
even amended laws to prevent the malicious sharing of fake media contents
generated by machine learning software like DeepFake. In order to determine the
authenticity of images and avoid the potential security issues, it is of
importance to identify DNG images.

Identification of fake images is much related to the research field of
information security, particularly in image forensics \cite{korus2017digital}
and biometric anti-spoofing \cite{galbally2014biometric}. Due to the fact that
fabricating a fake image would inevitably introduce some traces, the key to the
identification is to analyze and extract features that represent the
corresponding traces. For example, the quantization artifacts are used in JPEG
image forensics \cite{luo2010jpeg,bianchi2012image}, the joint artifacts left by
different image operations can be used to determine the operation chains
\cite{chen2017detection,zhang2019robustness,stamm2013forensically}, the splicing
inconsistences are exploited to locate tampered image regions
\cite{korus2016multi,li2017image}, and the displaying/imaging distortions are
utilized in face spoofing detection \cite{patel2016secure,boulkenafet2016face}.
{Recently, some works have been developed to identify fake images
generated by deep networks. Some of them employ the visual artifacts
\cite{matern2019exploiting} or saturation abnormalities
\cite{mccloskey2019detecting} in DNG images, and some of them are based on deep
learing paradigm
\cite{marra2018detection,mo2018fake,dang2018deep,zhuang2019detecting}.} Although
these works introduced some applicable approaches to identify DNG images, they
did not analyze the common inherent traces left in generated images and thus
failed to provide interpretable conclusions. Furthermore, they did not fully
consider some challenging scenarios, \eg, the image sources or the generative
models are different in the training and testing phases, which need to be
carefully coped with in practice. To this end, more efforts should be devoted to
the identification of DNG images.

{In this paper, we propose an effective and interpretable method to identify DNG images.
Considering the difference between camera imaging and DNG image generation, we
analyze the disparity between DNG images and real images and observe that the DNG images are more distinguishable in residual domain of chrominance components}.
Based on the observations,
we design a feature set to identify DNG images, which is composed by
co-occurrence matrix of residual images in different color components.
Furthermore, we address several challenging detection scenarios, including the
cases when the training and testing data are mismatched and the GAN model is unaware. {Extensive experimental results have shown the effectiveness of the proposed method: 1) When the training and testing data are matched, the 
proposed method achieves high accuracies. 2) When the training and testing images are generated by the same type of GAN but with different semantic content types, the proposed method outperforms existing methods; this situation has not been studied in previous works. 3) When the training and testing images are generated by different GANs, the proposed method also achieves superior performance, although the performance varies in different cases. 4) When the GAN model is unaware, the proposed method can obtain promising results with one-class classification.} The contributions of this paper are summarized as follows.
\begin{itemize}
  \setlength{\itemsep}{0pt}
  \item Considering that DNG image generation is different from camera imaging,
  we have conducted the analysis and demonstrated the disparities between DNG
  images and real images. {By measuring the correlations between
  adjacent pixels in some color spaces, we have found that the statistical
  properties of DNG images and real images are different in the chrominance
  components of HSV and YCbCr color spaces, while the disparities are more
  distinct in the residual domain.}

  \item We have proposed an effective feature set for DNG image identification.
  {The feature set consists of co-occurrence matrices extracted from 
  the residual images of several color components.} 
  The proposed feature set is of low dimension, and achieves
  good detection performance even with a small training set.

  \item We have evaluated the identification performance in different detection
  scenarios. The proposed method outperforms the existing methods when the image semantic types or the GAN models are mismatched in the training and testing phases. It is worthy to mention
  that, in the GAN model-unaware case (only real images are available for training),
  we can still achieve good results by performing one-class classification.
\end{itemize}

The rest of this paper is organized as follows. Section \ref{sec:related}
introduces some related works. Section \ref{sec:proposed} presents the details
of the proposed method. Section \ref{sec:experiment} reports and discusses the
experimental results. Finally, the concluding remarks are drawn in Section
\ref{sec:conclusion}.

\section{Related Works}
\label{sec:related}

\subsection{Image Generative Models and GANs}

A generative model can be trained with some given data, and it aims to produce
samples that follow the same distribution as the training data. In an ideal
case, by improving the model and increasing the amount and the quality of
training data, the generative model is expected to eventually generate plausible
samples similar to those coming from real world.
Currently the most popular generative models are based on deep neural networks,
including Generative Adversarial Networks (GANs)
\cite{goodfellow2014generative}, Variational Autoencoders (VAEs)
\cite{kingma2014auto}, and autoregressive models \cite{oord2016pixel}. VAEs and
autoregressive models usually produce images of poor quality compared to GANs,
which would limit their applications. Hence, most of the state-of-the-art image
generative models are built with GANs.

GAN was first proposed by Goodfellow \etal \cite{goodfellow2014generative}.
Basically, a GAN consists of two networks: a generator and a discriminator. The
generator tries to generate synthetic samples as the one drawing from real data
distribution, and the discriminator tries to correctly classify whether samples
are coming from the generator or the real data. The training of GAN works as
solving a two-player zero-sum game between the generator and the discriminator.
While the discriminator notices some differences between the real distribution
and the generated distribution, the generator adjusts its parameters to produce
samples closer to the real distribution. And then, the discriminator tries to
tell apart the two distributions again by adjusting its parameters. In an ideal
case, the generator is expected to eventually reproduce the distribution of real
data, and the discriminator fails to distinguish between generated samples and
real samples.
Up to now, many works
\cite{radford2016unsupervised,mao2017least,zhao2017energy,arjovsky2017wasserstein,
gulrajani2017improved,karras2018progressive,brock2018large,karras2018style} have
been proposed to improve the vanilla GAN. For example, Radford \etal
\cite{radford2016unsupervised} designed deep convolutional GAN (DCGAN), Arjovsky
\etal \cite{arjovsky2017wasserstein} adopted Wasserstein distance in GAN to make
the training more stable, Gulrajani \etal \cite{gulrajani2017improved} made an
improvement for Wasserstein GAN with gradient penalty (WGAN-GP). More recently,
the quality and variation of generated images have been further improved by
progressive growing of GANs (\textsc{Pro}GAN) \cite{karras2018progressive},
large scale training with architectural changes and modified regularization
scheme (\textsc{Big}GAN) \cite{brock2018large}, designing style-based generator
architecture (\textsc{Sty}GAN) \cite{karras2018style}, and generating images by
parts based on their conditional spatial coordinates (\textsc{Coco}GAN) \cite{lin2019coco}.

{In this paper, we consider DNG images generated from six popular GAN
models, including DCGAN \cite{radford2016unsupervised}, WGAN-GP
\cite{gulrajani2017improved}, \textsc{Pro}GAN \cite{karras2018progressive},
\textsc{Sty}GAN \cite{karras2018style}, \textsc{Big}GAN \cite{brock2018large},
and \textsc{Coco}GAN \cite{lin2019coco}. While some of these GANs have already
been considered in the recently-developed image database FaceForensic++
\cite{rossler2019faceforensics++}, we include some more different GAN models to
evaluate the generalization capability of detection methods.}

\subsection{Fake Image Identification}

Before the appearance of DNG images, many methods were developed to identify
fake images subjected to editing and/or rebroadcasting. Most of those methods
rely on specific traces regarding to the processing pipelines of fake images,
such as the JPEG compression errors of a recompressed image \cite{luo2010jpeg}
and the quality distortions of a spoofing face image
\cite{galbally2014face,wen2015face}. Hence, such targeted methods are not
suitable for detecting DNG images. {On the other hand, there are some
general methods that are based on either statistical features extracted
from image textures \cite{boulkenafet2016face,li2018identification} or deep
neural networks equiped with a constrained convolutional
layer\cite{bayar2018constrained}.} Such approaches may be used to detect DNG
images, as long as retraining a detector with DNG image samples. Nevertheless,
since these approaches do not take into account the traces left by the
generation pipeline of DNG images, they cannot always achieve satisfactory
performance.

With the growing interest in identifying DNG images, a few methods have been
designed to differentiate between DNG images and real images. Some of these
methods rely on feature engineering. McCloskey and Albright
\cite{mccloskey2019detecting} found that the normalization of an image generator
is different from a real camera, and proposed to detect GAN-generated images
with saturation cues. The best result of the area under the ROC curve (AUC)
obtained by this method was just around 0.70. Matern \etal
\cite{matern2019exploiting} reported that some generated face images exhibited
visual artifacts in eyes, teeth and facial contours, and utilized such artifacts
for identification, obtaining the AUC around 0.85. {Marra \etal
\cite{marra2019gans} showed that each GAN leaves a specific fingerprint in the
images generated by itself. Some low-level facial artifacts can be used to expose the images/videos generated by DeepFake, such as the absence of eye blinking \cite{li2018ictu}, the errors of estimated 3-D
head poses \cite{yang2019exposinga}, the inconsistences in facial parts
locations \cite{yang2019exposing}, and the correlations of facial expressions
and movements \cite{agarwal2019protecting}. However, such facial abnormalities
can be only applicable to generated face images/videos, and thus they are not suitable for detecting DNG images with other types of semantic contents.}

Other methods resort to deep learning (DL). Marra
\etal~\cite{marra2018detection} tested the performance of several DL
architectures for detecting images produced by image-to-image translation
\cite{zhu2017unpaired}. 
{
Mo \etal~\cite{mo2018fake} and Dang \etal
\cite{dang2018deep} designed customized convolutional neural networks (CNN) to
identify fake face images generated by GANs. Afchar
\etal~\cite{afchar2018mesonet} employed networks focusing on the mesoscopic
properties of images to perform detection. In addition to directly feeding the
images into CNNs, some works tried to improve the detection performance by
incorporating specific domain knowledge. For example, training the network with
co-occurrence matrices extracted from an image in the pixel domain of RGB space
\cite{nataraj2019detecting}, prompting the network to learning the affine face
warping artifacts \cite{li2019exposing} or up-sampling artifacts
\cite{zhang2019detecting}. }

{
A big potential
practical problem for identifying DNG images is the mismatch between the
training and testing data. In order to mitigate the performance degradation of a
trained detector in such case, researchers have tried to utilize more advanced
learning mechanisms, \eg, weakly-supervised transfer learning
\cite{cozzolino2018forensictransfer}, incremental learning
\cite{marra2019incremental} and two-step pairwise learning
\cite{zhuang2019detecting}. However, these approaches still need to
appropriately collect enough DNG images for the training procedure.
}


\section{DNG Image Identification}
\label{sec:proposed}

In this section, we first analyze some possible artifacts of DNG images and
investigate the disparities between DNG images and real images in
some color spaces. Then, we construct a feature set to capture the
artifacts of DNG images so as to identify them. Finally, we discuss several
detection scenarios and the corresponding detection strategies.

\subsection{Analysis from the Perspective of Color}
\label{subsec:analysis}

\subsubsection{The generation pipeline of DNG images}
In order to distinguish DNG images from real images, it needs to inspect the
artifacts left by GANs during creating images. Typically, the generator of a GAN
takes a random latent vector as input and employs several
convolution/deconvolution layers to gradually expand the spatial size of the
random vector. In the last layer of the generator, several feature maps are
transformed into a tensor with three channels, where the three channels
represent the R, G, and B components of the generated image, respectively.
{During this procedure, the convolution operations would introduce
some inherent properties into the DNG images. By contrast, a real image is
captured from real scene by camera, where the color components are decomposed
and digitalized from real world, meaning that the real pixels should be
inherently correlated in a different way.} Since the generation of DNG images is
quite different from camera
imaging, there should be certain disparities between DNG images and real images
from the perspective of color. Therefore, it is reasonable to assume that some
properties among the color components of DNG images are different from real
ones. In the following analysis, we will show some experimental evidences to
support this assumption.


\subsubsection{Discernibility of color component}
\label{subsubsec:discernibility}
As GANs usually generate images in RGB space, they tend to follow the properties
of real images in RGB space, while paying less attention to the properties in
other color spaces. In this way, although the differences between DNG images and
real ones are inapparent in RGB color space, they may be more obvious in other color
spaces. Therefore, we consider analyzing DNG images in three different color
spaces, \ie, RGB, HSV, and YCbCr, and try to use a metric to examine which color
component is more discernible for identifying DNG images. Firstly, we construct
the discernibility metric as follows.
\begin{enumerate}
  \item[a)]  For the $i$-th image $\mathbf{I}$ in a dataset, we calculate the
  correlation coefficient between the adjacent pixels in each of its color
  component $\mathbf{I}^c$ ($c\!\in\!\{\mathrm{R,G,B,H,S,V,Y,Cb,Cr}\}$), which can be formulated as:
  \begin{equation}\label{eq:corr2}
    r^c_i = \frac{\sum_{j = 1}^{m}\sum_{k = 1}^{n-1}\left(\mathbf{I}^c_{j,k}-\bar{\mathbf{I}}^c\right)\left(\mathbf{I}^c_{j,k+1}-\bar{\mathbf{I}}^c\right)}{\sqrt{\sum_{j = 1}^{m}\sum_{k = 1}^{n-1}\left(\mathbf{I}^c_{j,k}-\bar{\mathbf{I}}^c\right)^2\sum_{j = 1}^{m}\sum_{k = 1}^{n-1}\left(\mathbf{I}^c_{j,k+1}-\bar{\mathbf{I}}^c\right)^2} },
  \end{equation} 
  where $\bar{\mathbf{I}}^c$ is the mean value of $\mathbf{I}^c$, and $m$ and
  $n$ are the height and width of the image. In this way, the value of $r^c_i$
  represents the relevance of the adjacent pixel values. The larger the $r^c_i$,
  the higher correlation between the adjacent pixel values in $\mathbf{I}^c$.
  {Please note that we only employ the correlation analysis on horizontal
  adjacent pixels for simplifying the analysis, and similar results can be
  obtained by considering adjacent pixels in other directions.}
  



  \item[b)]
  For a set of DNG images, we calculate $r^c_i$ for each image and construct
  the histogram of $r^c_i$ as $\mathbb{H}_{\text{DNG}}^c$. Similarly, for
  a set of real images, we construct the histogram $\mathbb{H}_{\text{Real}}^c$.
  Then, we measure the similarity between the two histograms with Chi-square distance
  \begin{equation}\label{eq:chi_dist}
      d_{\chi^2}(\mathbb{H}_{\text{DNG}}^c,\mathbb{H}_{\text{Real}}^c) = \frac{1}{2}\sum_{x}\frac{\left(\mathbb{H}_{\text{DNG}}^c(x)-\mathbb{H}_{\text{Real}}^c(x)\right)^2}{\mathbb{H}_{\text{DNG}}^c(x)+\mathbb{H}_{\text{Real}}^c(x)},
  \end{equation}
  where $x$ is the bin index.
  $d_{\chi^2}(\mathbb{H}_{\text{DNG}}^c,\mathbb{H}_{\text{Real}}^c)$ can serve
  as the \emph{discernibility metric}. The larger the discernibility metric, the
  disparities between DNG images and real images are more significant. In other
  words, the color component $c$ is more powerful for distinguishing between DNG
  images and real images with a larger value of $d_{\chi^2}(\mathbb{H}_{\text{DNG}}^c,\mathbb{H}_{\text{Real}}^c)$ .
\end{enumerate}

\begin{figure*}[t]
  \centering
  \subfloat[R ($d_{\chi^2}$=0.002)]{\label{subfig:scorehist_R}
    \includegraphics[width=3.9cm]{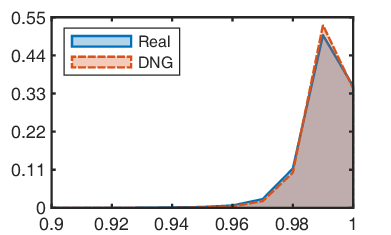}} 
  \subfloat[G ($d_{\chi^2}$=0.004)]{\label{subfig:scorehist_G}
    \includegraphics[width=3.9cm]{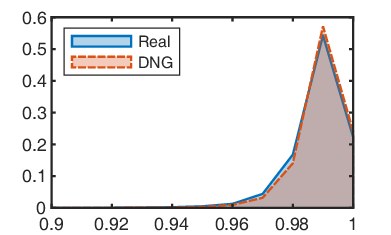}} 
  \subfloat[B ($d_{\chi^2}$=0.005)]{\label{subfig:scorehist_B}
    \includegraphics[width=3.9cm]{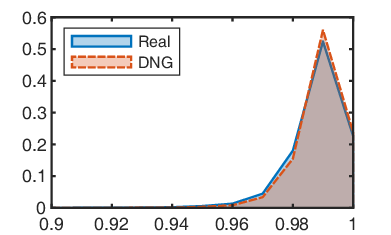}} \\ \vspace{-0.5em}
  \subfloat[H ($d_{\chi^2}$=0.019)]{\label{subfig:scorehist_H}
    \includegraphics[width=3.9cm]{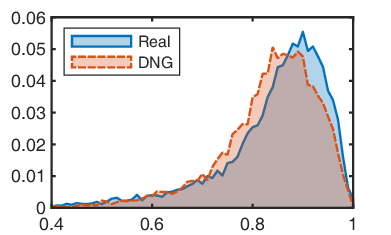}} 
  \subfloat[S ($d_{\chi^2}$=0.011)]{\label{subfig:scorehist_S}
    \includegraphics[width=3.9cm]{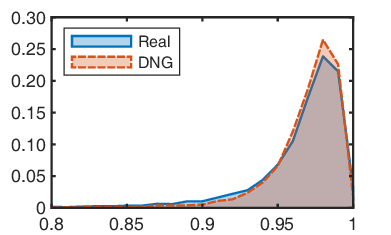}} 
  \subfloat[V ($d_{\chi^2}$=0.002)]{\label{subfig:scorehist_V}
    \includegraphics[width=3.9cm]{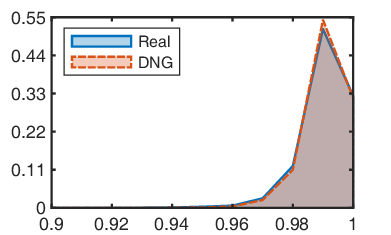}} \\ \vspace{-0.5em}
  \subfloat[Y ($d_{\chi^2}$=0.004)]{\label{subfig:scorehist_Y}
    \includegraphics[width=3.9cm]{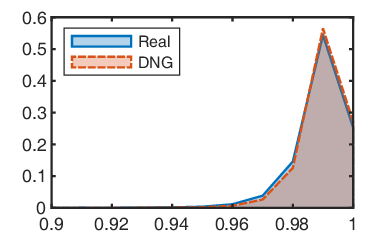}} 
  \subfloat[Cb ($d_{\chi^2}$=0.035)]{\label{subfig:scorehist_Cb}
    \includegraphics[width=3.9cm]{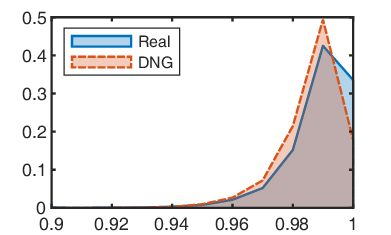}} 
  \subfloat[Cr ($d_{\chi^2}$=0.017)]{\label{subfig:scorehist_CR}
    \includegraphics[width=3.9cm]{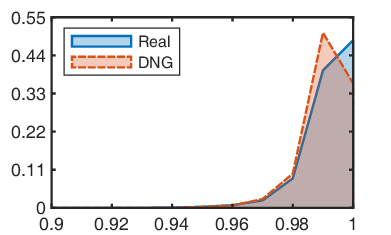}} \\ 
  \caption{The histograms $\mathbb{H}_{\text{DNG}}^c$ (red) and $\mathbb{H}_{\text{Real}}^c$ (blue) for different color components. The values of $d_{\chi^2}(\mathbb{H}_{\text{DNG}}^c ,\mathbb{H}_{\text{Real}}^c)$  are included in the sub-captions. } 
  \label{fig:scorehist}
\end{figure*}

To evaluate the effectiveness of the discernibility metric, we perform some
analytical experiments. We use the \textsc{Sty}GAN \cite{karras2018style} model
trained with the FFHQ dataset \cite{karras2018style} as an example of image
generative model. With the trained model, we generate a large amount of DNG
images. We randomly select 10,000 DNG images and 10,000 real images from the
generated dataset and the FFHQ dataset, respectively, and then compute the
$r^c_i$ ($c\!\in\!\{\mathrm{R,G,B,H,S,V,Y,Cb,Cr}\}$) as introduced above. Then,
we construct the histograms $\mathbb{H}_{\text{DNG}}^c$ and
$\mathbb{H}_{\text{Real}}^c$. The histograms for different color components are
shown in Fig. \ref{fig:scorehist}. {It can be observed that the
non-overlapping regions of $\mathbb{H}_{\text{DNG}}^c$ and
$\mathbb{H}_{\text{Real}}^c$ are quite small for the R, G, and B channels,
meaning that the disparities between DNG images and real images are not apparent
in the RGB color space. This can be explained by the DNG generation process
where R, G, B channels are directly generated to follow the distributions of real images. Via converting the images from RGB space into HSV/YCbCr
spaces, the R, G, and B channels are merged by linear or non-linear combinations
to form the other color channels. In this way, the disparities between DNG
images and real images will be amplified, and thus the non-overlapping regions
of $\mathbb{H}_{\text{DNG}}^c$ and $\mathbb{H}_{\text{Real}}^c$ become larger,
especially for the four chrominance components, \ie, H, S, Cb, and Cr. Such
phenomena can be supported by the discernibility metrics, \ie,
$d_{\chi^2}(\mathbb{H}_{\text{DNG}}^c,\mathbb{H}_{\text{Real}}^c)$. As shown in
the sub-captions of the corresponding sub-figures in Fig. \ref{fig:scorehist},
we observe that the values of $d_{\chi^2}(\mathbb{H}_{\text{DNG}}^c
,\mathbb{H}_{\text{Real}}^c)$ for the four chrominance components are all exceed
0.011, which are larger than those for the luminance components (\ie, V and Y)
and R, G, and B components (all being less than 0.005). These results indicate
that the chrominance components are more discernible than the others.}

\subsubsection{{Discernibility analysis in first-order differential residual domain}}

\begin{figure*}[t]
  \centering
  \subfloat[R ($d_{\chi^2}$=0.028)]{\label{subfig:scorehist_dR}
    \includegraphics[width=3.9cm]{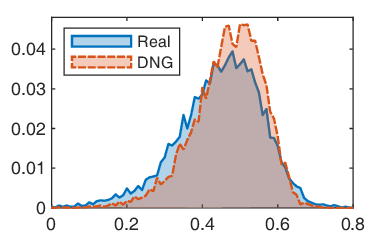}} 
  \subfloat[G ($d_{\chi^2}$=0.031)]{\label{subfig:scorehist_dG}
    \includegraphics[width=3.9cm]{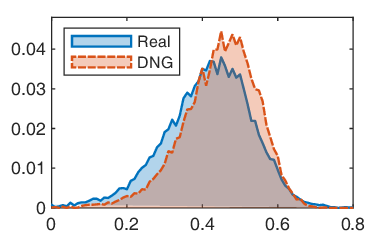}} 
  \subfloat[B ($d_{\chi^2}$=0.030)]{\label{subfig:scorehist_dB}
    \includegraphics[width=3.9cm]{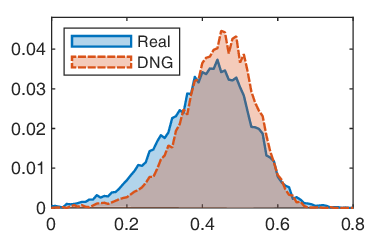}} \\ \vspace{-0.5em}
  \subfloat[H ($d_{\chi^2}$=0.040)]{\label{subfig:scorehist_dH}
    \includegraphics[width=3.9cm]{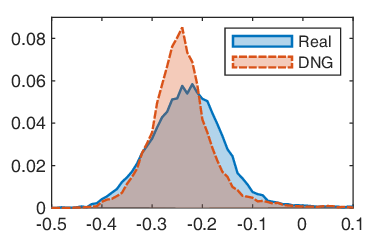}} 
  \subfloat[S ($d_{\chi^2}$=0.035)]{\label{subfig:scorehist_dS}
    \includegraphics[width=3.9cm]{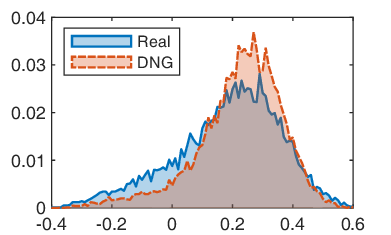}} 
  \subfloat[V ($d_{\chi^2}$=0.023)]{\label{subfig:scorehist_dV}
    \includegraphics[width=3.9cm]{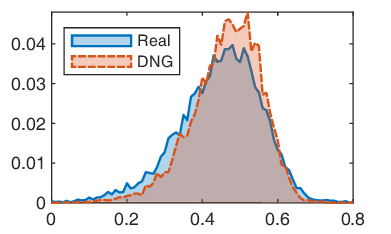}} \\ \vspace{-0.5em}
  \subfloat[Y ($d_{\chi^2}$=0.030)]{\label{subfig:scorehist_dY}
    \includegraphics[width=3.9cm]{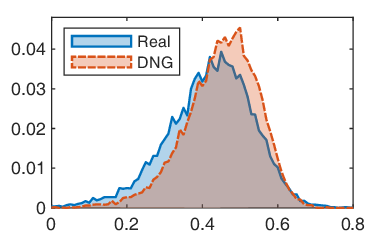}} 
  \subfloat[Cb ($d_{\chi^2}$=0.083)]{\label{subfig:scorehist_dCb}
    \includegraphics[width=3.9cm]{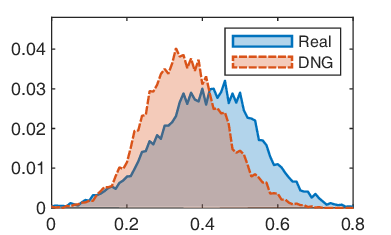}} 
  \subfloat[Cr ($d_{\chi^2}$=0.107)]{\label{subfig:scorehist_dCR}
    \includegraphics[width=3.9cm]{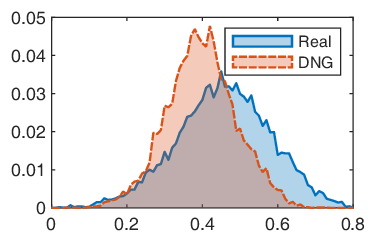}} \\ 
  {
  \caption{The histograms $\mathbb{H}_{\text{DNG}}^c$ (red) and
  $\mathbb{H}_{\text{Real}}^c$ (blue) for different color components in the
  residual domain. The
  values of $d_{\chi^2}(\mathbb{H}_{\text{DNG}}^c ,\mathbb{H}_{\text{Real}}^c)$
  are included in the sub-captions. }\label{fig:scorehist_diff}} 
\end{figure*}

{
In the above analysis, we directly derive the observations from the image
spatial domain. However, the contents of DNG images and real
images are not always visually distinguishable, meaning that the contents in DNG images and real images are quite similar. This would have negative impacts on analyzing their disparities. Therefore, it is more reasonable to
suppress image contents with high-pass filtering and then investigate the
disparities in image residuals. In fact, extracting features from high-pass
filtered image residuals have been successfully used in some imperceptible
pattern recognition applications, such as image steganalysis
\cite{fridrich2012rich} and image forensics \cite{li2018identification}.
}

{We apply a first-order differential operator as an example to obtain
image residuals, namely, 
\begin{equation}\label{eq:residual}
  \mathbf{R}^c_{j,k} = \mathbf{I}^c_{j,k}-\mathbf{I}^c_{j,k+1}, ~c\!\in\!\{\mathrm{R,G,B,H,S,V,Y,Cb,Cr}\}.
\end{equation}
where $\mathbf{I}^c$ is the $c$-th component of image $\mathbf{I}$ and $\mathbf{R}^c$ is the corresponding residual.
We only consider horizontal difference here, and similar results can be obtained
by considering vertical difference. Having the image residuals in residual
domain, we replace $\mathbf{I}^c$ and $\bar{\mathbf{I}}^c$ in Equation
(\ref{eq:corr2}) with $\mathbf{R}^c$ and $\bar{\mathbf{R}}^c$, respectively, and
then conduct the discernibility analysis with the same image datasets as
described in Section \ref{subsubsec:discernibility}. The histograms
$\mathbb{H}_{\text{DNG}}^c$ and $\mathbb{H}_{\text{Real}}^c$ for different color
components in residual domain are shown in Fig. \ref{fig:scorehist_diff}. From
this figure, it is observed that the non-overlapping regions of
$\mathbb{H}_{\text{DNG}}^c$ and $\mathbb{H}_{\text{Real}}^c$ become much more
noticeable, implying that the DNG image and real image are indeed more
distinguishable in the residual domain. This phenomenon can also be explained by
the fact that DNG images are undergone a different processing pipeline from real
images so that their statistics are different in the residual domain. Moreover,
the observations obtained in spatial domain still hold in residual domain: 1)
the non-overlapping regions of $\mathbb{H}_{\text{DNG}}^c$ and
$\mathbb{H}_{\text{Real}}^c$ for the H, S, Cb,Cr components are larger than those
for R, G, B, V, Y components, and 2) the discernibility metrics for the four
chrominance components are also greater than those for the other components. We
have also conducted the same analysis on images generated by other types of GANs
and obtained consistent results, indicating that some statistical features
extracted from the chrominance components in residual domain would be beneficial
for identifying DNG images.}

\begin{figure*}[t]
  \centering
  \includegraphics[scale=0.75]{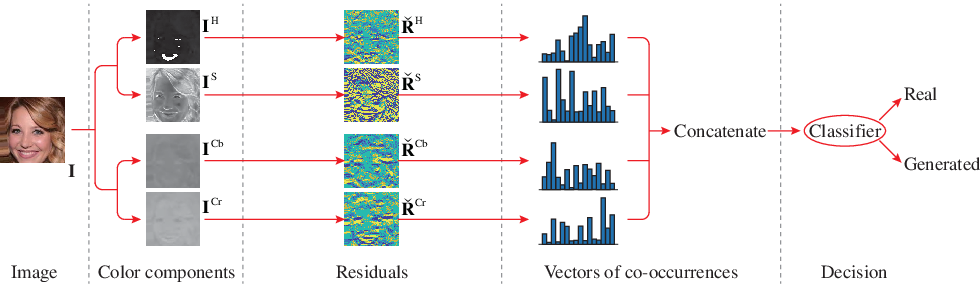}
  \caption{The overall framework of the proposed method.}
  \label{fig:framework}
\end{figure*}

\vspace{1em}
\subsection{Extracting Features from Color Components}
\label{subsec:fea}

{Based on the previous analysis, we decide to extract features from
the chrominance components in residual domain of an image. The overall framework
of the proposed method is illustrated in Fig. \ref{fig:framework}. For a given
image, we first compute the features from color components and then concatenate
them into a feature vector, and finally train a classifier to predict whether
the image is real or is generated by deep networks. In the feature extraction
stage, we first calculate the image residuals with the first-order differential
operators and pre-process them with truncation. Then, the co-occurrence matrix
\cite{haralick1973textural}, which has been widely used in image textural
analysis as a kind of feature descriptor, is extracted from the image residuals
and merged to form the feature set.}



\subsubsection{Truncating image residuals}
{Once the image residuals of H, S, Cb, Cr components have been obtained with
the first-order differential operators, we need to pre-process the residuals
before computing co-occurrence matrix. The reason is that there are too many
distinct element values in the residuals, and it would result in a co-occurrence
matrix with huge dimension if the raw residual data is directly used.} In order
to reduce the number of distinct values, the residual images
$\mathbf{R}^c~(c\!\in\!\{\mathrm{H,S,Cb,Cr}\})$ are truncated as follow:
\begin{equation}\label{eq:trucate}
 \check{\mathbf{R}}^c(x,y) = \begin{cases}
 \tau, & \mathbf{R}^c(x,y) \geq \tau, \\
 \mathbf{R}^c(x,y), & -\tau < \mathbf{R}^c(x,y) < \tau, \\
 -\tau, & \mathbf{R}^c(x,y) \leq -\tau,
 \end{cases}
\end{equation}
where $(x,y)$ is the position index of an element within the residual image and
$\tau\!>\!0$ is the truncation threshold. After truncation, the resulting residual
images $\check{\mathbf{R}}^c$ only contain integer values within the range of
$[-\tau,\tau]$. They are then used to compute the co-occurrence matrices.


\subsubsection{Extracting co-occurrence features}
In total, we have four co-occurrence matrices, which are calculated from
$\check{\mathbf{R}}^{\mathrm{H}}$, $\check{\mathbf{R}}^{\mathrm{S}}$, $\check{\mathbf{R}}^{\mathrm{Cb}}$,
and $\check{\mathbf{R}}^{\mathrm{Cr}}$, respectively. Typically, the co-occurrence matrix of a 2-D array
$\mathbf{V}$ is computed by
\begin{equation}\label{eq:comat}
\begin{split}
   &\mathbf{C}{(\theta_1,\theta_2,\ldots,\theta_d)} = \frac{1}{n}\sum_{x,y}\mathds{1} \Big( \mathbf{V}(x,y)=\theta_1,\\
   &\quad \quad \mathbf{V}(x+\Delta x,y+\Delta y)=\theta_2,\ldots, \\
     & \quad \quad \mathbf{V}({x+(d-1)\Delta x,y+(d-1)\Delta y})=\theta_d \Big),
\end{split}
\end{equation}
where $\mathds{1}(\cdot)$ is an indicator function, $(\theta_1,\theta_2,\ldots,\theta_d)$ is
the index of co-occurrence matrix, $d$ is the order of co-occurrence matrix, $n$
is the normalization factor, and $\Delta x$, $\Delta y$ are the offsets for two
neighboring elements. The dimensionality of each the co-occurrence matrix for
$\check{\mathbf{R}}^c$ is $(2\tau+1)^d$.

Since the co-occurrence matrix is symmetric, we can decrease the feature
dimension by combining the two bins, $\mathbf{C}{(\theta_1,\theta_2,\ldots,\theta_d)}$ and
$\mathbf{C}{(\theta_d,\theta_{d-1},\ldots,\theta_1)}$, into one bin. After combination, the
dimensionality of co-occurrence matrix is substantially decreased: the
co-occurrence matrices for
$\check{\mathbf{R}}^c~(c\!\in\!\{\mathrm{H,S,Cb,Cr}\})$ have only
$((2\tau+1)^d+(2\tau+1)^{d-1})/2$ bins.
Please note that the reduction of feature dimension is motivated by the natural
symmetry property of images, thus it would not significantly decrease the
detection performance. In fact, the resulting relatively low dimensional
features will speed up the training of classifier.


\subsubsection{Practical implementation}
{In our practical implementation, we use two first-order differential
operators, one in horizontal direction and one in vertical direction, to obtain
the image residuals.} The residuals are then processed as described above, where
the truncation threshold is set as $\tau\!=\!2$, and the order of co-occurrence
matrix is set as $d\!=\!3$. We choose the offsets as $(\Delta x$, $\Delta y) \in
\{(0,1), (1,0)\}$, and thus we have 4 co-occurrence matrices (2 residuals
$\times $ 2 offsets) for each color component. Finally, we take the element-wise
sum of the 4 co-occurrence matrices as the features. In total, a 300-D feature
set is obtained, in which the feature dimension for each of
$\check{\mathbf{R}}^{\mathrm{H}}$, $\check{\mathbf{R}}^{\mathrm{S}}$,
$\check{\mathbf{R}}^{\mathrm{Cb}}$, and $\check{\mathbf{R}}^{\mathrm{Cr}}$ is
$(5^3+5^2)/2=75$.

\subsection{Detection Scenarios and Strategies}
\label{subsec:detect_strategy}

In practice, there are many types of generative models, and they can be trained
with different real data for producing images with diverse contents. As a
result, DNG images generated by different models that are trained with different
datasets may not exhibit the same characteristic, leading to
difficulties in distinguishing them from real images. Based on the information
that is available in the training phase and the source of testing data, we
divide the detection scenarios into three cases, and discuss the corresponding
detection strategies as follows.

\subsubsection{Matched training-testing data}
In this case, the training and testing DNG images are generated by the same
model trained with the same real image dataset. {This is the simplest
case and was considered by default in some existing works, \eg,
\cite{marra2018detection,mo2018fake,mccloskey2019detecting,dang2018deep}}. To
perform the detection, the investigator just need to train a binary classifier
with real images and DNG images, and uses the trained classifier to predict the
class labels for the testing images.

\subsubsection{Mismatched training-testing data}
In this case, the training and testing DNG images are from different sources.
For example, they are generated by the same type of GAN but with different
datasets or even different image semantic content types, or they have the same
type of image semantic contents but are generated with different GANs. Binary
classification is also used in this case. Due to the mismatched sources, this
case can be exploited to evaluate the generalization ability of a detection
method. The better results a method achieves in this case, the better the method
is expected to perform in real applications. {Some existing works
(\eg,
\cite{cozzolino2018forensictransfer,marra2019incremental,zhuang2019detecting})
have been aware of this problem and tried to improve the performance with
advanced learning mechanisms, but they have not intensively studied the
performance for such mismatch cases.}

\subsubsection{Model-unaware case}
It is not rare that the investigator does not have any knowledge about the
generative model, and thus has no DNG image samples when building a classifier.
This is the most challenging case presented to the investigator. {To
our best knowledge, such situation has never been considered in the existing
works on DNG image identification.} To cope with this case, a possible way is to
train a one-class classifier with only real images and use the classifier to
detect whether the testing image is real or not. If a testing image is not
predicted as real, then it will be treated as a generated one.

\section{Experiments}
\label{sec:experiment}

In this section, we will evaluate the performance of the proposed method.
We first introduce the common experimental settings, and then present the
experimental results for the three detection scenarios described in Section
\ref{subsec:detect_strategy}.


\subsection{Experimental Setups}
\subsubsection{Real image datasets}
Five real images datasets with two types of semantic contents (\ie, face and bedroom) and
different resolutions were used in the experiments. Each real image dataset is
denoted as $\mathcal{R}_\alpha$, where the subscript $\alpha$ indicates 
image source and resolution. The details of the real image datasets
are described as follows, and their basic information can be referred to the
upper part of Table \ref{tbl:datasets}.
\begin{itemize}
  \setlength{\itemsep}{0pt}
  \item \emph{Low-Resolution (LR) Face} ($\mathcal{R}_{\text{F-LR}}$ and
    $\mathcal{R}_{\text{Fl-LR}}$): The $\mathcal{R}_{\text{F-LR}}$ dataset
    consists of 200,000 celebrity face images that were randomly selected from
    the ``Align\&Cropped'' PNG images in the CelebA dataset \cite{liu2015deep}.
    We first cropped a $138\times 138$ facial region from each image to remove
    the background, and then resized the cropped region to $128\times 128$. The
    $\mathcal{R}_{\text{Fl-LR}}$ dataset contains 10,000 face images in the LFW
    dataset \cite{huang2007labeled}. These images are with the original size of
    $250\times 250$. For each image, we first cropped a $150\times 150$ facial
    region and then resize it to $128\times 128$. Please note that since
    $\mathcal{R}_{\text{Fl-LR}}$ contains much fewer images than others, the
    images in this dataset (and the DNG images related to this datasets) are
    only used for testing in the data-mismatch case (Section
    \ref{subsubsec:mismatched-sub1}).

  \item \emph{Low-Resolution Bedroom} ($\mathcal{R}_{\text{B-LR}}$): This
    dataset consists of 200,000 bedroom images that were randomly selected from
    the LSUN Bedroom dataset \cite{yu2015lsun}. We first cropped the central
    region with the size of  $256\times 256$ from each image, and then resized
    the cropped region into $128\times 128$.
  
  \item \emph{High-Resolution (HR) Face} ($\mathcal{R}_{\text{F-HR}}$): This
    dataset consists of 100,000 face images, including 30,000 images in the
    CelebA-HQ dataset \cite{karras2018progressive} and 70,000 images in the FFHQ
    dataset \cite{karras2018style}. {These images are with the size of
    $1024\times 1024$, and they are resized to $256\times 256$ in experiments so
    as to reduce the computational time (especially for deep learning based
    methods). Although the images have been pre-processed with down-scaling, the
    detection decision obtained by the classifier can also be applied to the
    original large images.}
  
  \item \emph{High-Resolution Bedroom} ($\mathcal{R}_{\text{B-HR}}$): This
    dataset consists of 100,000 bedroom images, which were randomly selected
    from the LSUN Bedroom dataset \cite{yu2015lsun}. We cropped the central
    $256\times 256$ region from each image.
\end{itemize}

\begin{table*}
  \centering
  \caption{The real and generated datasets used in the experiments.}\label{tbl:datasets}
  \begin{threeparttable}
  \scalebox{0.75}{
  \begin{tabular}{*5cl} \toprule
    Dataset                            & Category & Content & Resolution     & Quantity & \multicolumn{1}{c}{Note} \\ \midrule 
    $\mathcal{R}_{\text{F-LR}}$  & Real     & Face    & $128\times 128$ & 200,000  & Selected from CelebA \\ 
    $\mathcal{R}_{\text{Fl-LR}}$  & Real     & Face    & $128\times 128$ & \phantom{0}10,000  & Selected from LFW \\ 
    $\mathcal{R}_{\text{B-LR}}$   & Real     & Bedroom & $128\times 128$ & 200,000  & Selected from LSUN bedroom \\
    $\mathcal{R}_{\text{F-HR}}$   & Real     & Face    &{$1024\times 1024$}& 100,000  & Combination of CelebA-HQ and FFHQ \\
    $\mathcal{R}_{\text{B-HR}}$   & Real     & Bedroom & $256\times 256$ & 100,000  & Selected from LSUN bedroom \\ \midrule
    $\mathcal{G}^*_{\text{F-LR}}$\tnote{$\dagger$} & Generated& Face    & $128\times 128$ & 200,000  & GANs trained with CelebA \\ 
    $\mathcal{G}^*_{\text{Fl-LR}}$ & Generated& Face    & $128\times 128$ & \phantom{0}10,000  & GANs trained with LFW \\ 
    $\mathcal{G}^*_{\text{B-LR}}$ & Generated& Bedroom & $128\times 128$ & 200,000  & GANs trained with LSUN bedroom \\
    $\mathcal{G}^*_{\text{F-HR}}$ & Generated& Face    &{$1024\times 1024$}& 100,000
    & GANs trained with CelebA-HQ or FFHQ \\ 
    $\mathcal{G}^*_{\text{B-HR}}$ & Generated& Bedroom & $256\times 256$ &
    100,000  & GANs trained with bedroom images \\ \bottomrule
  \end{tabular}}
  \begin{tablenotes}
    \scriptsize
    \item[$\dagger$] The asterisk $*$ denotes the type of GAN.
    $*\in\{\text{DCGAN},\text{WGAN-GP}\}$ for LR datasets, \\
    {and $*\in\{\text{\textsc{Pro}GAN},\text{\textsc{Sty}GAN},\text{\textsc{Big}GAN},\text{\textsc{Coco}GAN}\}$
    for HR bedroom datasets, \\
    and $*\in\{\text{\textsc{Pro}GAN},\text{\textsc{Sty}GAN}\}$ for HR face
    datasets.}
  \end{tablenotes}
  \end{threeparttable}
\end{table*}

\subsubsection{DNG image datasets}
{In the experiments, we adopted DNG images generated by six types of GANs,
including DCGAN \cite{radford2016unsupervised}, WGAN-GP
\cite{gulrajani2017improved}, \textsc{Pro}GAN \cite{karras2018progressive}, 
\textsc{Sty}GAN \cite{karras2018style}, \textsc{Big}GAN
\cite{brock2018large}, and \textsc{Coco}GAN \cite{lin2019coco}.} For DCGAN and WGAN-GP, we
used the
implementation available online\footnote{Available at
\url{https://www.github.com/igul222/improved_wgan_training}.} and trained GANs
with the LR real image datasets (\ie, $\mathcal{R}_{\text{F-LR}}$, $\mathcal{R}_{\text{Fl-LR}}$, and
$\mathcal{R}_{\text{B-LR}}$). Hence, the trained GANs produce LR DNG images
with the size of $128\times 128$. For \textsc{Pro}GAN, \textsc{Sty}GAN,
\textsc{Big}GAN and \textsc{Coco}GAN, we
used the corresponding trained models\footnote{Available at
\url{https://drive.google.com/open?id=0B4qLcYyJmiz0NHFULTdYc05lX0U},
\url{http://stylegan.xyz/drive}, \url{https://tfhub.dev/deepmind/biggan-128/2},
and
\url{https://drive.google.com/drive/folders/1r-BvW6cVMHKJw-0wMI6mUepMkboWwWqN},
respectively. 
} released by the authors to generate HR images (\ie,
$1024\times 1024$ face images and/or $256\times 256$ bedroom images). The
$1024\times 1024$ face images were resized to $256\times 256$ in our
experiments. For simplicity, in the following context we denote a DNG images
dataset as $\mathcal{G}_\alpha^\beta$, where $\alpha$ represents image
source and resolution, and $\beta$ represents the type of GAN model. For
example, $\mathcal{G}^{\text{DCGAN}}_{\text{F-LR}}$ denotes the LR CelebA face images
generated by a DCGAN model. The basic information of the DNG image datasets are
summarized in the bottom part of Table \ref{tbl:datasets}. 
{In total, more than 1,400,000 DNG images are involved in our experiments.}

\subsubsection{Comparative study}
{We have compared the proposed method with four hand-crafted feature
based methods and five deep learning based methods, including general-purpose
methods and targeted methods.} The details are as follows.
\begin{itemize}
  \setlength{\itemsep}{0pt}
  \item SRM \cite{fridrich2012rich} ({General-purpose hand-crafted
  feature}, 34671-D): It is originally designed for image steganalysis. The
  features are extracted from image residuals obtained by a series of high-pass
  filters.

  \item Sub-SRM \cite{li2018identification} ({General-purpose
  hand-crafted feature}, 714-D): This is a refined subset of SRM. It achieved
  good performance in image forensics.

  \item CoALBP+LPQ \cite{boulkenafet2016face} ({General-purpose
  hand-crafted feature}, 19968-D): This method is designed for face spoofing
  detection. The feature set is composed of Co-Occurrence of Adjacent Local
  Binary Patterns and Local Phase Quantization.

  \item Sat-Cues \cite{mccloskey2019detecting} ({Targeted hand-crafted
  feature for DNG image detection}, 24-D): It uses the saturation cues to
  classify the real images and images generated by GANs.

  \item VGG-16 \cite{simonyan2014very} ({General-purpose DL based
  method}): A famous CNN for image classification.

  \item ResNet\_v2-50 \cite{he2016identity} ({General-purpose DL based
  method}): Another famous CNN for image classification, which uses identity
  connections to improve performance.

  \item Mo \etal~\cite{mo2018fake} ({Targeted DL based method for DNG
  image detection}): A CNN for detecting the images generated by
  \textsc{Pro}GAN.

  \item {CGFace \cite{dang2018deep} (Targeted DL based method for DNG
   image detection): A customized CNN for computer-generated face detection.}
  
  \item {TS-CDNN \cite{zhuang2019detecting} (Targeted DL based method
  for DNG image detection): A coupled deep neural network (CDNN) with a two-step
  learning approach for detecting GAN generated face images. }
\end{itemize}
Please note that we accordingly modified the fully connected layers of the
original CNN networks if necessary, so as to ensure that they are compatible
with the image sizes.


\subsubsection{Training and testing protocol}
Unless otherwise specified, we followed the descriptions below for training and
testing. In each experiment, 50,000 pairs of DNG images and real images (25\% of
the LR images and 50\% of the HR images) were
randomly selected for training, while the remaining images were used for
testing. For the hand-crafted feature based methods, we trained the
ensembles of LDA (linear discriminative analysis) learners
\cite{kodovsky2012ensemble} as classifiers. The parameters were set as defaults
used in \cite{kodovsky2012ensemble}. For the DL based methods, we trained
the networks with 90\% of the training data and kept 10\% for validation. {The
Momentum optimizer was used in training the network of Mo \etal~and VGG-16,
while the Adam optimizer was used in training ResNet\_v2-50, CGFace, and TS-CDNN;} the learning rates
were searched within a certain range and the optimal ones were respectively
selected for the networks. {The networks were trained 50 epochs with a batch size
of 64 (for the network of Mo \etal~, VGG-16, ResNet\_v2-50, and CGFace) or 128
(for TS-CDNN)}. We saved the models every 1 epoch and chose the ones with the best
validation accuracies as the final models. In the testing stage, we computed and
reported the false positive rate (FPR, the probability that real images are
identified as generated ones), false negative rate (FNR, the probability that
generated images are identified as real ones), and the overall accuracy (ACC).

\subsection{Performance on Matched Training-testing Data}
\label{subsec:matched}

\begin{table*}[t]
  \centering
  {
  \caption{Classification results (\%) for \textbf{matched} training and testing data.}\label{tbl:matched}}
  \scalebox{0.85}{
  \begin{tabular}{lrrr} \toprule
    \multirow{2}{*}{Method} & \multicolumn{1}{c}{FPR} & \multicolumn{1}{c}{FNR} &
    \multicolumn{1}{c}{ACC} \\
     & Average (Best/Worst) & Average (Best/Worst) & Average (Best/Worst) \\ \midrule
     Sub-SRM \cite{li2018identification}    & \phantom{0}0.11 (\phantom{0}0.00 / \phantom{0}0.82) & \phantom{0}0.05 (\phantom{0}0.00 / \phantom{0}0.40) & 99.92 (100.0 / 99.39) \\
     SRM \cite{fridrich2012rich}            & \phantom{0}0.03 (\phantom{0}0.00 / \phantom{0}0.19) & \phantom{0}0.01 (\phantom{0}0.00 / \phantom{0}0.07) & 99.98 (100.0 / 99.87) \\
     CoALBP+LPQ \cite{boulkenafet2016face}  & \phantom{0}0.01 (\phantom{0}0.00 / \phantom{0}0.08) & \phantom{0}0.01 (\phantom{0}0.00 / \phantom{0}0.03) & \textbf{99.99} (100.0 / 99.95) \\
     Sat-Cues \cite{mccloskey2019detecting} & 30.66 (17.77 / 42.76) & 25.53 (17.59 / 38.59) & 71.91 (80.69 / 59.33) \\
     VGG-16 \cite{simonyan2014very}         & \phantom{0}0.28 (\phantom{0}0.01 / \phantom{0}1.32) & \phantom{0}0.55 (\phantom{0}0.01 / \phantom{0}2.40) & 99.58 (99.99 / 98.14) \\
     ResNet\_v2-50 \cite{he2016identity}    & \phantom{0}0.56 (\phantom{0}0.01 / \phantom{0}1.99) & \phantom{0}0.74 (\phantom{0}0.01 / \phantom{0}2.57) & 99.35 (99.99 / 98.04) \\
     Mo \etal \cite{mo2018fake}             & \phantom{0}0.51 (\phantom{0}0.00 / \phantom{0}3.51) & \phantom{0}0.56 (\phantom{0}0.00 / \phantom{0}2.43) & 99.46 (100.0 / 97.03) \\
     CGFace \cite{dang2018deep}             & \phantom{0}4.75 (\phantom{0}0.04 / 37.81) & \phantom{0}4.67 (\phantom{0}0.01 / 41.78) & 95.29 (99.97 / 60.20) \\
     TS-CDNN \cite{zhuang2019detecting}     & \phantom{0}7.04 (\phantom{0}0.04 / 22.32) & 10.47 (\phantom{0}0.20 / 67.30) & 91.24 (99.88 / 63.91) \\
     Proposed                               & \phantom{0}0.25 (\phantom{0}0.00 / \phantom{0}1.96) & \phantom{0}0.27 (\phantom{0}0.00 / \phantom{0}2.14) & 99.74 (100.0 / 97.95) \\
    \bottomrule
  \end{tabular}}
\end{table*}

In this subsection, we evaluate the performance for identifying DNG images in
the case that the training and testing data are matched. {For each
detection method, we respectively trained 10 classifiers to classify the fake images in the 10 DNG image
datasets (\ie, $\mathcal{G}_{\text{F-LR}}^{\text{DCGAN}}$,
$\mathcal{G}_{\text{F-LR}}^{\text{WGAN-GP}}$,
$\mathcal{G}_{\text{B-LR}}^{\text{DCGAN}}$, 
$\mathcal{G}_{\text{B-LR}}^{\text{WGAN-GP}}$, 
$\mathcal{G}_{\text{F-HR}}^{\text{\textsc{Pro}GAN}}$,
$\mathcal{G}_{\text{F-HR}}^{\text{\textsc{Sty}GAN}}$,
$\mathcal{G}_{\text{B-HR}}^{\text{\textsc{Pro}GAN}}$,
$\mathcal{G}_{\text{B-HR}}^{\text{\textsc{Sty}GAN}}$, 
$\mathcal{G}_{\text{B-HR}}^{\text{\textsc{Big}GAN}}$, and
$\mathcal{G}_{\text{B-HR}}^{\text{\textsc{Coco}GAN}}$)
and their corresponding real images. The results are
summarized in Table \ref{tbl:matched}, where the average, the best, and the
worst results of FPR, FNR, and ACC obtained by the 10 classifiers for each method are reported. From Table \ref{tbl:matched}, it
is observed that all methods achieve an average accuracy over 90\% except the
method based on saturation cues.}
These results indicate
that the DNG images can be accurately identified when a classifier is trained
with corresponding data. Among all the methods, CoALBP+LPQ achieves the best
average performance (99.99\%), and the proposed method obtains competitive
results (only 0.25\% less). Please note that comparing with the high-dimensional
CoALBP+LPQ feature, the proposed feature set is of much lower dimension, which
can significantly reduce computational cost. 

\begin{figure}[t]
  \centering
  \includegraphics[scale=0.6]{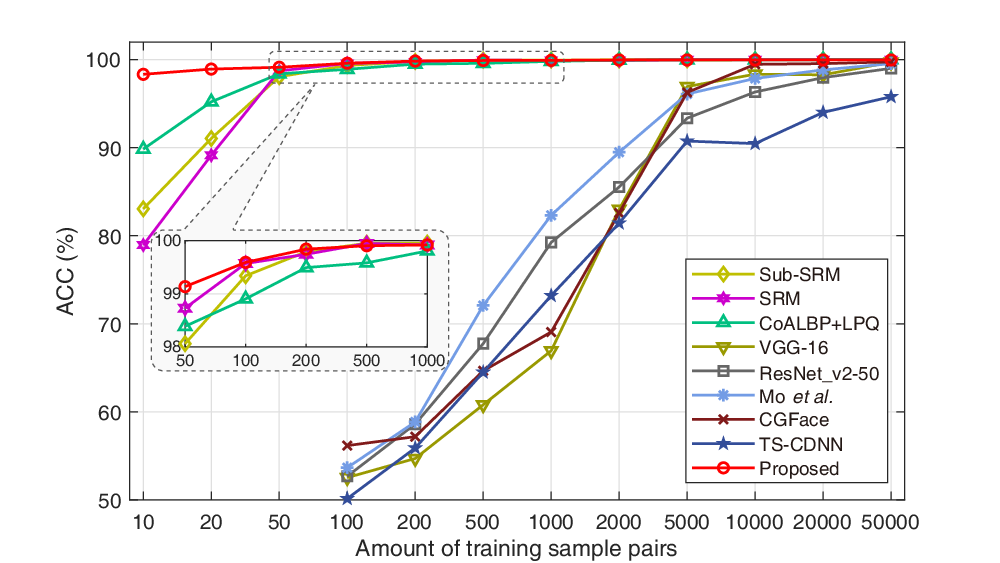}
  \caption{Classification accuracies for $\left\{\mathcal{R}_{\text{F-LR}},\mathcal{G}_{\text{F-LR}}^{\text{WGAN-GP}}\right\}$ with different amounts of training sample pairs.}
  \label{fig:trnnum}
\end{figure}

Another merit of the proposed feature is its advantages for small amount of training samples. We conducted an extra
experiment to classify the images in $\mathcal{R}_{\text{F-LR}}$ and
$\mathcal{G}_{\text{F-LR}}^{\text{WGAN-GP}}$ by reducing training data. Fig.
\ref{fig:trnnum} shows the testing accuracies obtained with different amounts of
training image pairs. It is observed that the proposed method achieves the
testing accuracy of 98\% even when using 10 pairs of training images,
significantly outperforming the others in the same case. It means that
the proposed method is less dependent on the amount of training data. When the
number of training image pairs reaches 1,000, Sub-SRM, SRM, and CoALBP+LPQ can
obtain almost the same performance with the proposed method, while the DL based
methods only obtain accuracies that are less than 83\%. According to Fig.
\ref{fig:trnnum}, The DL based methods require more than 20,000 training image
pairs to get competitive performance with the proposed method. 

\subsection{Performance on Mismatched Training-testing Data}
\label{subsec:mismatched}

In practice, the image source or the type of GAN model may not be fully known by
the investigator. In this case, there is a mismatch between the training and
testing data. In this subsection, we consider three sub-cases. In the first case,
the training and testing images are from different sources and have the
same type of semantic content, such as face or bedroom, while in the second case the image semantic content types are
different in the training and testing phases. The last case is that the types of
GAN models are mismatched.

\subsubsection{Mismatched image sources (same semantic type)}
\label{subsubsec:mismatched-sub1}

\begin{table*}[t]
  \centering
  {
  \caption{Classification results (\%) for \textbf{mismatched image sources} (same semantic type).}
  \label{tbl:dataset-mismatched-face}}
  \scalebox{0.85}{
  \begin{tabular}{L{3.0cm}L{3.5cm}rrr} \toprule
    & Method    & FPR & FNR & ACC \\ \midrule
    \multirow{8}{3.0cm}{\footnotesize{\emph{train:}} \\ $\left\{\mathcal{R}_{\text{F-LR}},\mathcal{G}_{\text{F-LR}}^{\text{DCGAN}}\right\}$ \\ \footnotesize{\emph{test:}} \\ $\left\{\mathcal{R}_{\text{Fl-LR}},\mathcal{G}_{\text{Fl-LR}}^{\text{DCGAN}}\right\}$}
    & Sub-SRM \cite{li2018identification}    & 0.00 & 0.00 & \textbf{100.0} \\
    & SRM \cite{fridrich2012rich}            & 0.00 & 0.00 & \textbf{100.0} \\
    & CoALBP+LPQ \cite{boulkenafet2016face}  & 0.00 & 0.00 & \textbf{100.0} \\
    & Sat-Cues \cite{mccloskey2019detecting} & 42.68 & 10.65 & 73.34 \\
    & VGG-16 \cite{simonyan2014very}         & 1.86 & 94.98 & 51.58 \\
    & ResNet\_v2-50 \cite{he2016identity}    & 0.00 & 100.0 & 50.00 \\
    & Mo \etal \cite{mo2018fake}             & 0.00 & 0.00 & \textbf{100.0} \\
    & CGFace \cite{dang2018deep}             & 1.25 & 99.02 & 49.87 \\
    & TS-CDNN \cite{zhuang2019detecting}     & 0.04 & 99.85 & 50.06 \\
    & Proposed                               & 0.00 & 0.00 & \textbf{100.0} \\
    \midrule[.03em]
    \multirow{8}{3.0cm}{\footnotesize{\emph{train:}} \\ $\left\{\mathcal{R}_{\text{F-LR}},\mathcal{G}_{\text{F-LR}}^{\text{WGAN-GP}}\right\}$ \\ \footnotesize{\emph{test:}} \\ $\left\{\mathcal{R}_{\text{Fl-LR}},\mathcal{G}_{\text{Fl-LR}}^{\text{WGAN-GP}}\right\}$}
    & Sub-SRM \cite{li2018identification}    & 0.01 & 0.00 & \textbf{100.0} \\
    & SRM \cite{fridrich2012rich}            & 0.05 & 0.00 & 99.98 \\
    & CoALBP+LPQ \cite{boulkenafet2016face}  & 0.00 & 0.11 & 99.95 \\
    & Sat-Cues \cite{mccloskey2019detecting} & 65.54 & 25.06 & 54.70 \\
    & VGG-16 \cite{simonyan2014very}         & 1.87 & 0.45 & 98.84 \\
    & ResNet\_v2-50 \cite{he2016identity}    & 8.29 & 87.98 & 51.87 \\
    & Mo \etal \cite{mo2018fake}             & 52.73 & 19.59 & 63.84 \\
    & CGFace \cite{dang2018deep}             & 0.64 & 1.30 & 99.03 \\
    & TS-CDNN \cite{zhuang2019detecting}     & 14.83 & 39.76 & 72.71 \\
    & Proposed                               & 0.00 & 2.50 & 98.75 \\
    \bottomrule
  \end{tabular}}
\end{table*}

It is reasonable to assume that the architecture of GAN is known, but the real
images used to train the model are not available. To perform the detection, the
investigator first trains a GAN with the same network architecture by using an
alternative dataset, and then use the trained GAN to produce DNG samples. With
these samples, the investigator can train a binary classifier to detect DNG
images. In this experiment, we trained the binary classifiers with the LR face
images related to the CelebA dataset (\ie, $\mathcal{R}_{\text{F-LR}}$,
$\mathcal{G}_{\text{F-LR}}^{\text{DCGAN}}$, and
$\mathcal{G}_{\text{F-LR}}^{\text{WGAN-GP}}$), and than tested them with the LR
face images related to the LFW dataset (\ie, $\mathcal{R}_{\text{Fl-LR}}$,
$\mathcal{G}_{\text{Fl-LR}}^{\text{DCGAN}}$, and
$\mathcal{G}_{\text{Fl-LR}}^{\text{WGAN-GP}}$). The detection results are shown
in Table \ref{tbl:dataset-mismatched-face}. It can be observed that all
hand-crafted features except Sat-Cues achieve high accuracies in both cases,
while the DL based methods perform poor in either of the two cases (VGG-16, Mo
\etal, CGFace, and TS-CDNN), or both (ResNet\_v2-50). The proposed method can perfectly detect the
images generated by DCGAN even when they are from different sources, and also
obtains very good result for detecting the images generated by WGAN-GP, though
underperforms Sub-SRM by 1.25\%.

\subsubsection{Mismatched image sources (different semantic types)}

\begin{table*}[t]
  \centering
  {
  \caption{Classification results (\%\!) for \textbf{mismatched image sources} (different semantic types).}
  \label{tbl:dataset-mismatched}}
  \scalebox{0.85}{
  \begin{tabular}{lrrr} \toprule
    \multirow{2}{*}{Method} & \multicolumn{1}{c}{FPR} & \multicolumn{1}{c}{FNR} &
    \multicolumn{1}{c}{ACC} \\
     & Average (Best/Worst) & Average (Best/Worst) & Average (Best/Worst) \\ \midrule
     Sub-SRM \cite{li2018identification}    & 23.72 (\phantom{0}0.00 / 99.96) & 17.54 (\phantom{0}0.00 / 84.88) & 79.37 (99.94 / 50.02) \\
     SRM \cite{fridrich2012rich}            & 23.04 (\phantom{0}0.00 / 99.58) & 17.77 (\phantom{0}0.00 / 99.00) & 79.59 (100.0 / 50.21) \\
     CoALBP+LPQ \cite{boulkenafet2016face}  & 28.08 (\phantom{0}0.00 / 98.77) & 12.86 (\phantom{0}0.00 / 88.28) & 79.53 (99.99 / 49.74) \\
     Sat-Cues \cite{mccloskey2019detecting} & 42.72 (16.11 / 62.53) & 45.54 (16.64 / 78.97) & 55.87 (64.15 / 50.46) \\
     VGG-16 \cite{simonyan2014very}         & 19.81 (\phantom{0}0.01 / 79.47) & 61.26 (\phantom{0}3.74 / 99.98) & 59.46 (91.05 / 49.99) \\
     ResNet\_v2-50 \cite{he2016identity}    & \phantom{0}4.30 (\phantom{0}0.00 / 23.40) & 95.19 (75.28 / 99.99) & 50.26 (51.34 / 49.64) \\
     Mo \etal \cite{mo2018fake}             & 17.45 (\phantom{0}0.02 / 87.48) & 50.01 (\phantom{0}0.08 / 99.99) & 66.27 (99.95 / 48.48) \\
     CGFace \cite{dang2018deep}             & 16.77 (\phantom{0}0.07 / 44.16) & 63.87 (\phantom{0}0.82 / 99.85) & 59.68 (81.97 / 45.96) \\
     TS-CDNN \cite{zhuang2019detecting}     & 11.71 (\phantom{0}0.01 / 28.37) & 79.40 (49.92 / 99.97) & 54.45 (72.13 / 50.02) \\
     Proposed                               & 28.33 (\phantom{0}0.00 / 96.29) & \phantom{0}9.45 (\phantom{0}0.00 / 41.53) & \textbf{81.11} (99.99 / 51.85) \\
    \bottomrule
  \end{tabular}}
\end{table*}

To simulate the mismatch of image semantic contents, we first used the classifiers
trained with face images to classify the bedroom images, and then used the
classifiers trained with bedroom images to classify the face images. 
{
For example, a classifier trained with $\mathcal{R}_{\text{F-LR}}$ and
$\mathcal{G}_{\text{F-LR}}^{\text{DCGAN}}$ is used to classify the images in
$\mathcal{R}_{\text{B-LR}}$ and $\mathcal{G}_{\text{B-LR}}^{\text{DCGAN}}$, and
another classifier trained with $\mathcal{R}_{\text{B-LR}}$ and
$\mathcal{G}_{\text{B-LR}}^{\text{DCGAN}}$ is used to classify the images in
$\mathcal{R}_{\text{F-LR}}$ and $\mathcal{G}_{\text{F-LR}}^{\text{DCGAN}}$. The
testing for LR images and HR images were conducted separately, and the DNG
images in training and testing stages are produced by the same type of GAN, so
there are totally 8 cases (Please note that the DNG images generated by
\textsc{Big}GAN and \textsc{Coco}GAN were not included in this experiment since
they are all bedroom images).}
The average, best,
and worst results are shown in Table \ref{tbl:dataset-mismatched}.
We observe that the DL based methods significantly underperform most of the
feature based methods, implying that they over-fit on image contents and thus
have poor generalization performance for images with different semantic types.
The proposed method obtains lower FNR than others and achieves the best
average performance (81.11\%), outperforming the second place (\ie, SRM)
by 1.5\%. {On the other hand, it is observed that the detection
performance varies greatly in different cases. For example, the proposed method
achieves 99\% accuracy on the best case (testing $\mathcal{R}_{\text{B-LR}}$ and
$\mathcal{G}_{\text{B-LR}}^{\text{DCGAN}}$ with the classifier trained on $\mathcal{R}_{\text{F-LR}}$ and
$\mathcal{G}_{\text{F-LR}}^{\text{DCGAN}}$), while it behaves close to random
guessing in the worst case (testing $\mathcal{R}_{\text{F-HR}}$ and
$\mathcal{G}_{\text{F-HR}}^{\text{\textsc{Pro}GAN}}$ with the classifier trained on $\mathcal{R}_{\text{B-HR}}$ and
$\mathcal{G}_{\text{B-HR}}^{\text{\textsc{Pro}GAN}}$).}

\subsubsection{Mismatched GAN models}

\begin{table*}[t]
  \centering
  {
  \caption{Classification results (\%) for \textbf{mismatched GAN models} in training and testing.}
  \label{tbl:GAN-mismatched}}
  \scalebox{0.85}{
  \begin{tabular}{lrrr} \toprule
    \multirow{2}{*}{Method} & \multicolumn{1}{c}{FPR} & \multicolumn{1}{c}{FNR} &
    \multicolumn{1}{c}{ACC} \\
     & Average (Best/Worst) & Average (Best/Worst) & Average (Best/Worst) \\ \midrule
     Sub-SRM \cite{li2018identification}    & \phantom{0}0.06 (\phantom{0}0.00 / \phantom{0}0.82) & 39.45 (\phantom{0}0.00 / 100.0) & 80.25 (100.0 / 50.00) \\
     SRM \cite{fridrich2012rich}            & \phantom{0}0.02 (\phantom{0}0.00 / \phantom{0}0.19) & 35.11 (\phantom{0}0.00 / 100.0) & 82.44 (100.0 / 50.00) \\
     CoALBP+LPQ \cite{boulkenafet2016face}  & \phantom{0}0.01 (\phantom{0}0.00 / \phantom{0}0.08) & 22.05 (\phantom{0}0.00 / 100.0) & 88.97 (100.0 / 49.99) \\
     Sat-Cues \cite{mccloskey2019detecting} & 26.79 (17.77 / 42.76) & 40.18 (17.19 / 63.48) & 66.52 (78.43 / 48.82) \\
     VGG-16 \cite{simonyan2014very}         & \phantom{0}0.34 (\phantom{0}0.01 / \phantom{0}1.32) & 77.09 (\phantom{0}0.35 / 99.94) & 61.29 (99.77 / 49.92) \\
     ResNet\_v2-50 \cite{he2016identity}    & \phantom{0}0.54 (\phantom{0}0.01 / \phantom{0}1.99) & 82.09 (12.82 / 100.0) & 58.68 (93.56 / 49.64) \\
     Mo \etal \cite{mo2018fake}             & \phantom{0}0.32 (\phantom{0}0.00 / \phantom{0}3.51) & 58.04 (\phantom{0}0.00 / 99.99) & 70.82 (99.95 / 48.39) \\
     CGFace \cite{dang2018deep}             & \phantom{0}6.87 (\phantom{0}0.04 / 37.81) & 71.18 (\phantom{0}0.20 / 99.91) & 60.98 (99.84 / 48.54) \\
     TS-CDNN \cite{zhuang2019detecting}     & \phantom{0}6.96 (\phantom{0}0.04 / 22.32) & 84.19 (55.68 / 99.96) & 54.42 (68.16 / 46.81) \\
     Proposed                               & \phantom{0}0.14 (\phantom{0}0.00 / \phantom{0}1.96) & 16.12 (\phantom{0}0.00 / 100.0) & \textbf{91.87} (100.0 / 50.00)\\
    \bottomrule
  \end{tabular}}
\end{table*}

To simulate the mismatch of GAN models, we used the classifiers trained with the
images generated by a certain GAN to classify the images generated by another
GAN. The experiments for LR images and HR images were also conducted separately,
{and thus there are totally 18 cases (2 for LR face images, 2 for LR
bedroom images, 2 for HR face images, and 12 for HR bedroom images). The
experimental results are shown in Table \ref{tbl:GAN-mismatched}. On average,
the proposed method performs the best (91.87\%)  among all the methods, since it
obtains good accuracies for testing the HR bedroom images even when the GAN
models are mismatched. However, it dose not always have good performance. In the
worst case, namely, using the classifier trained with
$\mathcal{R}_{\text{F-LR}}$ and $\mathcal{G}_{\text{F-LR}}^{\text{DCGAN}}$ to
test the images in $\mathcal{R}_{\text{F-LR}}$ and
$\mathcal{G}_{\text{F-LR}}^{\text{WGAN-GP}}$, all the DNG images are mistakenly
classified, resulting in the accuracy of 50\%. The reason may be that DCGAN is
simpler than WGAN-GP, and thus the images generated by DCGAN are more
distinguishable than those generated by WGAN-GP, leading to that the trained
classifier is too simple to identify the images generated by WGAN-GP. This
indicates that the DNG samples must be carefully selected during training a
binary classifier, otherwise the trained detector would produce wrong decisions
in some mismatch cases.}





\subsection{Performance on Model-unaware One-class Classification}
\label{subsec:one-class}

{As mentioned above, the performance of a binary classifier for
identifying DNG images may vary when the training and testing data are mismatch,
and thus it is important to train the binary classifier with compatible DNG images.
However, in practice the GAN model is sometimes unknown and there are no
suitable DNG images available for training, thus it is difficult to build a
feasible binary classifier.} To perform the detection, an alternative way is to
build a one-class classifier, which fits a model via learning the properties of
real images and regard the DNG images as outliers. In this subsection, we
evaluate the performance of the proposed method in one-class classification.
Since it is not a trivial task to adapt a deep network for one-class
classification, we only considered the feature based methods in this experiment
(Sat-Cues was also excluded for its poor performance). We employed the support
vector machine (SVM) with Gaussian kernel implemented in LIBSVM
\cite{chang2011libsvm} as one-class classifier. The parameter $\gamma$ of the
Gaussian kernel was determined via a grid search. The parameter $\upsilon$,
which controls the upper bound of training error (\ie, the probability of that
regarding the training real images as outliers), was set as 0.10 and 0.05,
respectively. {In this experiment, the real face images and real
bedroom images are combined together, and thus there are a LR real image
dataset $\mathcal{R}_{\text{F-LR}}+\mathcal{R}_{\text{B-LR}}$ and a HR real
image dataset $\mathcal{R}_{\text{F-HR}}+\mathcal{R}_{\text{B-HR}}$}. Hence, we
trained four classifiers for each method, corresponding to
the two values of $\upsilon$ and LR/HR images, respectively. In the training
phase, 50,000 real images were randomly selected to train the one-class SVMs. In
the testing phase, the remaining real images and all DNG images were fed
to the trained models to calculate the accuracies.

\begin{table*}[t]
  \centering
  {
  \caption{Detection accuracies (\%) obtained by \textbf{one-class classification}.}\label{tbl:one-class}}
  \begin{threeparttable}
  \scalebox{0.85}{
  \begin{tabular}{cC{1.5cm}l*4{C{1.3cm}}} \toprule
    \multirow{2}{*}{$\upsilon$} & \multirow{2}{=}{\centering Train dataset} & 
    \multirow{2}{*}{Method} & 
    \multirow{2}{=}{\centering
    $\mathcal{R}_{\text{F-LR}}$\\\vspace{-0.6em}{\scriptsize
    +}\\\vspace{-0.6em}$\mathcal{R}_{\text{B-LR}}$} & 
    \multirow{2}{=}{\centering Best} & 
    \multirow{2}{=}{\centering Worst} & 
    \multirow{2}{*}{Average} \\
    \\ 
    \midrule
    \multirow{4}{*}{0.10} 
    & \multirow{4}{=}{\centering $\mathcal{R}_{\text{F-LR}}$\\+\\$\mathcal{R}_{\text{B-LR}}$}
    &  Sub-SRM \cite{li2018identification}  & 88.99 & 100.0 & 12.56 & 65.94 \\
    & 
    &  SRM \cite{fridrich2012rich} & 89.81 & 100.0 & 2.15 & 53.32 \\
    & 
    & CoALBP+LPQ \cite{boulkenafet2016face} & 89.64 & 100.0 & 13.49 & 67.80 \\
    & 
    & Proposed   & 89.90 & 100.0 & 99.85 & \textbf{99.94} \\
    \midrule[0.03em]
    \multirow{4}{*}{0.05} 
    & \multirow{4}{=}{\centering $\mathcal{R}_{\text{F-LR}}$\\+\\$\mathcal{R}_{\text{B-LR}}$}
    &  Sub-SRM \cite{li2018identification}    & 93.76 & 100.0 & 8.00 & 62.82 \\
    & 
    &  SRM \cite{fridrich2012rich} & 94.93 & 100.0 & 0.55 & 51.51 \\
    & 
    & CoALBP+LPQ \cite{boulkenafet2016face} & 94.66 & 100.0 & 4.35 & 59.45 \\
    & 
    & Proposed   & 94.91 & 100.0 & 99.19 & \textbf{99.69} \\
    \bottomrule
    \toprule
    \multirow{2}{*}{$\upsilon$} & \multirow{2}{=}{\centering Train dataset} & 
    \multirow{2}{*}{Method} & 
    \multirow{2}{=}{\centering
    $\mathcal{R}_{\text{F-HR}}$\\\vspace{-0.6em}{\scriptsize
    +}\\\vspace{-0.6em}$\mathcal{R}_{\text{B-HR}}$} & 
    \multirow{2}{=}{\centering Best} & 
    \multirow{2}{=}{\centering Worst} & 
    \multirow{2}{*}{Average} \\
    \\
    \midrule
    \multirow{4}{*}{0.10} 
    & \multirow{4}{=}{\centering $\mathcal{R}_{\text{F-HR}}$\\+\\$\mathcal{R}_{\text{B-HR}}$}
    &  Sub-SRM \cite{li2018identification}   & 89.65 & 99.95 & 4.44 & 46.38 \\
    & 
    &  SRM \cite{fridrich2012rich} & 90.15 & 77.44 & 5.67 & 30.71 \\
    & 
    & CoALBP+LPQ \cite{boulkenafet2016face} & 90.04 & 67.62 & 1.33 & 27.04 \\
    & 
    & Proposed   & 89.86 & 98.09 & 4.89 & \textbf{49.82} \\
    \midrule[0.03em]
    \multirow{4}{*}{0.05} 
    & \multirow{4}{=}{\centering $\mathcal{R}_{\text{F-HR}}$\\+\\$\mathcal{R}_{\text{B-HR}}$}
    &  Sub-SRM \cite{li2018identification}   & 94.29 & 99.81 & 1.92 & \textbf{41.26} \\
    & 
    &  SRM \cite{fridrich2012rich} & 95.03 & 64.14 & 2.56 & 22.66 \\
    & 
    & CoALBP+LPQ \cite{boulkenafet2016face} & 95.20 & 50.51 & 0.45 & 17.60 \\
    & 
    & Proposed   & 94.78 & 94.57 & 2.02 & 39.49 \\
    \bottomrule
  \end{tabular}}
  \begin{tablenotes}
    \scriptsize
    \item[{$\dagger$}] {$\mathcal{R}_{\text{F-LR}}+\mathcal{R}_{\text{B-LR}}$ denotes the combination of LR real image datasets $\mathcal{R}_{\text{F-LR}}$ and $\mathcal{R}_{\text{B-LR}}$, \\ and $\mathcal{R}_{\text{F-HR}}+\mathcal{R}_{\text{B-HR}}$ denotes the combination of HR real image datasets $\mathcal{R}_{\text{F-HR}}$ and $\mathcal{R}_{\text{B-HR}}$.}
  \end{tablenotes}
  \end{threeparttable}
\end{table*}

The classification results are shown in Table \ref{tbl:one-class},
{where the top half part contains the results for LR images and the
bottom half part contains the results for HR images.} In this table,  we
report the accuracies for detecting real images, as well as the average, best,
and worst results for detecting DNG images. On the one hand, we observe that the
detection performance for real images is related to the parameter $\upsilon$.
Specifically, when $\upsilon$ is 0.10 or 0.05, the testing errors for real
images are correspondingly around 10\% or 5\%, respectively. It means that the
trained classifier can adequately model the distribution of real images. On the
other hand, it is observed that the proposed method can accurately detect the LR
DNG images, outperforming the other methods by a large margin. {For
the HR DNG images, although our method obtains poor result in the worst case
(for detecting the HR face images generated by \textsc{Pro}GAN), it
achieves relatively good performance on average.} Based on these
experimental results, it is promising to employ one-class classification with
the proposed features to identify DNG images when only real images are available
for training.


\section{Conclusion}
\label{sec:conclusion}

In this paper, we have investigated some new issues on identifying deep network
generated images. {We have analyzed and observed that the disparities
between DNG images and real images are apparent in the residual domain of the
chrominance color components.} Based on the observations, a feature set based on
color statistical features is proposed. The feature set is compact and
effective. {We have evaluated the performance in some situations with different assumption on the availability of training data.}
The experimental results show that the proposed features equipped with a binary
classifier can accurately differentiate between DNG images and real images when
the training and testing data are matched, {and outperforms existing methods when the image semantic types or the GAN models are mismatched in the training and testing phases}. Moreover, in the model-unaware case, the proposed
features can still effectively identify some types of DNG images with a
one-class classifier.

In addition to the proposed detection method, this paper also provides some
useful insights for the research community. Typically, the generative models
generate images by imitating real images in RGB color space. Although the
generated image may be visually visually appealing and be indistinguishable from real images for human eyes, they can be easily
detected by the proposed method. It means that many inherent properties of real
images, such as the properties in different color components, have not been
properly depicted by the existing generative models. In order to further improve
the quality of DNG images, more constrains should be considered in generative
models.

In the future, we will improve the performance of our method to meet the
requirements in practical detection situations. Furthermore, we will try to
detect the modern image processing techniques that utilize deep networks based
generative models as backbones, for example, image inpainting with GANs.

\section*{Acknowledgements}
This work was supported in part by NSFC (61802262, 61872244 and U19B2022),
Guangdong Basic and Applied Basic Research Foundation (2019B151502001),
Guangdong R\&D Program in Key Areas (2019B010139003), and Shenzhen R\&D
Program (JCYJ20180305124325555 and GJHZ20180928155814437).

\bibliographystyle{elsarticle-num}
\bibliography{ref}

\end{document}